\documentstyle[12pt]{article}

\def\Q{{\cal Q}}
\def\F{{\cal F}}

\begin{document}
\hbox to \hsize{\hfil DTP 95/5}
\hbox to \hsize{\hfil February, 1995}
\medskip
\centerline{{\Large\bf The Hamiltonian structure of the}}

\vspace{.2in}
\centerline{{\Large\bf dispersionless Toda hierarchy}}

\vspace{.2in}
\centerline{{\bf D.B.Fairlie}}

\vspace{.1in}
\centerline{Dept. of Mathematical Sciences, University of Durham,}
\vspace{.1in}
\centerline{Durham, DH1 3LE, England
\footnote{e-mail: david.fairlie@durham.ac.uk}.}

\vspace{.2in}
\centerline{{\bf I.A.B.Strachan}}

\vspace{.1in}
\centerline{Dept. of Mathematics and Statistics, University of Newcastle,}
\vspace{.1in}
\centerline{Newcastle-upon-Tyne, NE1 7RU, England
\footnote{e-mail: i.a.b.strachan@newcastle.ac.uk}.}

\vspace{.3in}
\centerline{{\bf Abstract}}

\vspace{.2in}
\small
\parbox{5.8in}{The Hamiltonian structure of the two-dimensional dispersionless
Toda hierarchy is studied, this being a particular example of a system of
hydrodynamic type. The polynomial conservation laws for the system turn
out, after a change of variable, to be associated with the axially symmetric
solutions of the 3-dimensional Laplace equation and this
enables a generating function for the Hamiltonian densities to be derived
in closed form.}
\normalsize

\bigskip
\section*{1. Introduction }

\bigskip

In this paper the Hamiltonian structure of the integrable system

\begin{equation}
\begin{array}{rcl}
S(x,t)_t & = & P(x,t)_x \,, \\
P(x,t)_t & = & P(x,t) S(x,t)_x \,,
\end{array}
\label{eq:SPtoda}
\end{equation}

\noindent (or equivalently, the equation $(\log P)_{tt} = P_{xx}\,$)
will be investigated. The reason for the interest in this system is
two-fold. Firstly, it appears in various physical applications, for example:

\smallskip

$\bullet$ In the construction of self-dual Vacuum and Einstein-Weyl metrics
\cite{Ward}.

\smallskip

$\bullet$ In topological field theory, as a solution to the WDVV equations
\cite{Krichever}.

\smallskip

$\bullet$ As the long-wave (or dispersionless) limit of the Toda equation
\cite{Saveliev,Hoppe}.

\smallskip

\noindent Secondly, despite its simplicity, it has the same generic properties
as more complicated systems,  so it can be used as a \lq test bed\rq~for
various ideas. (See for example the recent paper of
Mineev-Weinstein\cite{min}.) It is one of the simplest examples of an equation
of hydrodynamic
type that is, an equation of the form

\begin{equation}
u^i_t = V^i_j ({\bf u}) u^j_x \label{eq:hydro}
\end{equation}

\noindent about which there is an extensive literature
\cite{DubrovinNovikov,Tsarev}. Much of this, however,
involves general theory, and there are comparatively few specific examples. The
purpose of this paper is to derive concrete results for the dispersionless Toda
system. Rather than study (\ref{eq:SPtoda})~directly, a transformation will be
made to a new set $\{u,v\}$ of variables in which the equations are symmetric
under the interchange of $u$ and $v\,,$ this enabling results from the theory
of
homogeneous symmetric functions to be used. Explicitly, let $\{S,P\}$ be a
solution to (\ref{eq:SPtoda})~and let $\{u,v\}$ be the roots of the quadratic

\[
z^2 - S z + P = 0 \,.
\]

\noindent In these new variables the original system (\ref{eq:SPtoda})
becomes

\begin{equation}
\begin{array}{rcl}
u_t & = & u v_x \,, \\
v_t & = & v u_x \,,
\end{array}
\label{eq:UVtoda}
\end{equation}

\noindent Note that if $u=v$ then this reduces to the dispersionless
Korteweg-deVries equation.

\bigskip

This change of variables may be used to see the following property of any
solution
to (\ref{eq:UVtoda}).
Given $\{u,v\}$ satisfying (\ref{eq:UVtoda})~a new solution is given by the
roots of the quadratic

\[
z^2 - (u + v + k) z + uv = 0 \,.
\]

\noindent Moreover, one has a version of the \lq theorem of
permutability\rq~--applying this transformation first with constant $k_1$ then
with $k_2$ results in the same solution as applying the transformation with
constant $k_2$ followed by that with $k_1\,.$ The general solution to
(\ref{eq:SPtoda})~or
(\ref{eq:UVtoda})~may be found implicitly by performing a hodographic
transformation. Specific examples have been calculated in \cite{Kodama} and
the general form of the solution in \cite{Saveliev}.

\bigskip

Another motivation comes from the construction of integrable, relativistic
system in $(1+1)$ dimensions. Any solution to (\ref{eq:UVtoda}) of the form

\begin{eqnarray*}
u & = & { f \Big( {\partial\phi\over\partial t} {\partial\phi\over\partial x}
\Big) \over \Big( {\partial\phi\over\partial x} \Big)^2 } \,, \\
v & = & { g \Big( {\partial\phi\over\partial t} {\partial\phi\over\partial x}
\Big) \over \Big( {\partial\phi\over\partial x} \Big)^2 } \,,
\end{eqnarray*}

\noindent (or, equivalently, a similarity solution) will result in such a
system for $\phi\,,$ with $x$ and $t$ being light-cone variables. The equations
can be reduce to the single first order differential equation

\[
{dY\over dX} = {Y (1 + 2 X - X Y  )\over X (1 + 2 Y - X Y)}\,,
\]

\noindent this being an Abel equation of the second kind. Despite the
integrability of the original system (\ref{eq:UVtoda}) we have not been able to
solve this equation, except in the case $Y=X$ (or $u=v$) which results in the
well-known Born-Infeld equation for $\phi\,$ \cite{FairlieMulvey}.

\bigskip

The rest of this paper is arranged as follows. In section $2$ conservation laws
will be derived for this system, together with a generating function for them,
and in section $3$ the associated hierarchy will be studied. In section $4$
these
results will be used to study the Hamiltonian structure of this system and its
hierarchy. These results follow from direct calculation and do not use the
corresponding Lax pair. This Lax pair has a very simple solution, details of
which may be found in the appendix. Various generalisations are presented in
section
$5$.

\bigskip

\section*{2. Conservation Laws}

\bigskip

By direct calculation it is easy to find the first few polynomial conservation
laws:

\begin{eqnarray*}
{[ u+v ]_t} & = & [uv]_x \,, \\
{[ u^2 + 4 u v + v^2 ]_t} & = & [ 2 u v ( u + v ) ]_x \,, \\
{[ u^3 + 9 u v (u + v ) + v^3 ]_t} & = & [ 3 u v ( u^2 + 3 u v + v^2 ) ]_x \,,
\end{eqnarray*}

\noindent and a general such law will be denoted by

\begin{equation}
\Q_{n,t} = \F_{n,x} \,, \label{eq:conservation}
\end{equation}

\noindent where $\Q_n$ (the charge density) is an $n^{\rm th}$-order symmetric
polynomial in $u$ and $v\,,$ and $\F_n$ (the flux density) is an
$(n+1)^{\rm th}$-order symmetric polynomial in $u$ and $v\,.$ The coefficients
in
$\Q_n$ are the squares of the binomial coefficients $B(n,r)$:

\[
\Q_n = \sum_{r=0}^n B(n,r)^2 u^r v^{n-r}
\]

\noindent and

\[
\F_n = \sum_{r=0}^n \frac{n-r}{r+1} B(n,r)^2 u^{r+1} v^{n-r}\,.
\]

\noindent To prove this one assumes a general form for the symmetric polynomial
$\Q$ and adjust the coefficients so that the $t$-derivative is a total
$x$-derivative. These functions also satisfy the equations

\begin{equation}
\begin{array}{rcl}
{\F_{n,u}} & = & v {\Q_{n,v}}\,, \\
{\F_{n,v}} & = & u {\Q_{n,u}}\,,
\end{array}
\label{eq:one}
\end{equation}

\noindent from which follows the conservation law (\ref{eq:conservation}).
The hierarchy of conservation laws may be combined into a single generating
function

\[
\Q(\lambda)_t = \F(\lambda)_x \,,
\]

\noindent so that the coefficient of $\lambda^n$ in its formal power series
expansion is the conservation law (\ref{eq:conservation}). To show this it is
first necessary to find a recursion relation amongst the above charge
densities.

\bigskip

\indent From the fact that $\Q_n$ and $\F_n$ are homogeneous polynomials in $u$
and $v$
of degree $n$ and $(n+1)$ respectively one obtains the following relations

\begin{equation}
\begin{array}{rcl}
n \Q_n & = & u{\Q_{n,u}}+v{\Q_{n,v}}
\,, \\
(n+1) \F_n & = & u{\F_{n,u}}+v{\F_{n,v}}
\,
\end{array}
\label{eq:two}
\end{equation}

\noindent and

\begin{equation}
\begin{array}{rcl}
{\F_{n+1,u}} & = & (n+1) (\F_n + v \Q_n ) \,, \\
{\F_{n+1,v}} & = & (n+1) (\F_n + u \Q_n ) \,.
\end{array}
\label{eq:three}
\end{equation}

\noindent Using these, together with (\ref{eq:one}), one easily obtains the
following recursion relations:

\begin{eqnarray*}
\Q_{n+1} & = & 2 \F_n + (u+v) \Q_n \,,  \\
(n+2)\F_{n+1} & = & (n+1) \Big[ (u+v) \F_n + 2 uv\Q_n \Big]
\end{eqnarray*}

\noindent which results in the second order recurrence relation for the charge
densities:

\[
(n+2) \Q_{n+2} - (2n+3) (u+v) \Q_{n+1} + (n+1) (u-v)^2 \Q_n = 0\,.
\]

\noindent This bears a strong resemblance to the recursion relation for
Legendre polynomials. In fact, on defining

\[
\Q_n = (u-v)^n P_n \Bigg( {u+v\over u-v}  \Bigg)
\]

\noindent one obtains the Legendre recursion relation:

\[
(n+1) P_{n+1}(z) - (2n+1) z P_n(z) + n P_{n-1}(z) = 0
\]

\noindent and the first two conservations laws give $P_1(z)=1$ and
$P_2(z)=(3z^2-1)/2\,,$ the first two Legendre polynomials. Thus the
charge densities may be expressed in terms of Legendre polynomials. Using the
well-known generating function for these polynomials results in a generating
function for the conservation laws:

\[
{\partial{\phantom{t}}\over\partial t}
\Bigg[ {1\over \sqrt{ (u-v)^2\lambda^2 - 2(u+v)\lambda +1}} \Bigg] =
{\partial{\phantom{x}}\over\partial x}
\Bigg[ {\lambda(u+v) -1 \over \sqrt{ (u-v)^2\lambda^2-2(u+v)\lambda + 1}}
\Bigg]\,.
\]

\bigskip

The form of this results suggest that one should define a new set of variables

\begin{eqnarray*}
r & = & u - v \,, \\
\cos\theta & = & {u+v\over u-v}\,.
\end{eqnarray*}

\noindent In this systems the equation $(u\Q_u)_u=(v\Q_v)_v$ for the charge
density becomes the axially symmetric Laplace equation in three dimensions.
In terms of the inverse transformation

\begin{eqnarray*}
u(x,t) & = & - r(x,t) \cos^2 {\theta(x,t)\over 2}\,, \\
v(x,t) & = & \phantom{-} r(x,t) \sin^2 {\theta(x,t)\over 2}
\end{eqnarray*}

\noindent the original system (\ref{eq:UVtoda}) becomes

\begin{equation}
\begin{array}{rcl}
r_t & = & - (r^2/2)\,\, z_x \,, \\
z_t & = & \phantom{-} (1-z^2)/2\,\, r_x\,,
\end{array}
\label{eq:RZtoda}
\end{equation}

\noindent where $z=\cos\theta\,.$  The form of this suggest various
generalizations, which will be investigated in section $5\,.$

\bigskip

It is clear that  any solution to axially symmetric Laplace
equation, besides  the class of solutions solutions $r^n Q_n(\cos\theta)$ will
yield conservation laws. In particular one also has three other basic solution,
namely

\[
r^{-n-1} P_n(\cos\theta)\,,\hskip 20mm r^{-n-1} Q_n(\cos\theta)\,,\hskip 20mm
r^n Q_n(\cos\theta)\,,
\]

\noindent where $Q_n(z)$ is the Legendre function of the second kind. If one
wishes for conservation laws which remain regular as $r\rightarrow 0$ (i.e in
the limit in which the dispersionless Toda system becomes the
dispersionless KdV
equation) then this excludes the first two. However the third remains and gives
a family of non-polynomial conservation laws for (\ref{eq:UVtoda}), the first
few being

\begin{eqnarray*}
\Bigg[ {1\over 2} \log({u\over v}) \Bigg]_t & = & \Bigg[ {1\over 2} (v-u)
\Bigg]_x \,, \\
\Bigg[ {1\over 2} \Q_1 \log({u\over v}) - (u-v) \Bigg]_t & = & \Bigg[ {1\over
2} \F_1
\log({u\over v}) + {1\over 4} (v-u)(v+u) \Bigg]_x \,, \\
\Bigg[ {1\over 2} \Q_2 \log({u\over v}) - {3\over 2} (u^2 - v^2) \Bigg]_t & = &
\Bigg[ {1\over 2} \F_2\log({u\over v})+{1\over 6} (v-u)(v^2+10uv+v^2) \Bigg]_x
\end{eqnarray*}

\noindent (the logarithmic terms in the $(n+1)^{\rm th}$ such conservation law
are of the form ${1\over 2} \Q_n \log(u/v) \,$ and
${1\over 2} \F_n \log(u/v)\,$). Note that these are all
antisymmetric under the interchange $u\leftrightarrow v \,,$ and so vanish in
the limit $r\rightarrow 0\,.$

\bigskip

\section*{3. Generalised symmetries and commuting flows}

\bigskip

One fundamental property of an integrable system is the existence of an
infinite family of commuting flows. For the system (\ref{eq:UVtoda}) it is
straightforward to show that any such flow must be given by the system

\begin{eqnarray*}
u_{t'} & = & \F u_x + u \Q v_x \,, \\
v_{t'} & = & \F v_x + v \Q u_x \,
\end{eqnarray*}

\noindent where $\F$ and $\Q$ are any function of $u$ and $v$ (not necessarily
symmetric) which satisfy the relations (\ref{eq:one}). Thus for each of the
polynomial conservation laws derived in section $2$ one has an associated flow
labelled by the integer $n$ ($=$ degree of $\Q_n\,$),

\begin{equation}
\begin{array}{rcl}
u_{t_n} & = & \F_n u_x + u \Q_n v_x \,, \\
v_{t_n} & = & \F_n v_x + v \Q_n u_x \,.
\end{array}
\label{eq:hierarchy}
\end{equation}

\noindent It is convenient to let $t_0 = t \,, \F_0 = 0$ and $\Q_0=1\,,$ so
the original system corresponds to the $t_0$-flow. It follows from the general
theory of generalised symmetries that the function $\F u_x + u \Q v_x$ and
$\F v_x + v \Q u_x$ are characteristics for the original system, and that all
these flows commute \cite{Olver}.

\bigskip

The charges are conserved with respect to the $t_0$-flow, and one might expect
them also to be conserved with respect to the $t_n$-flow. This is
indeed the case:

\bigskip

\noindent {\bf Lemma}
\begin{equation}
{\partial\Q_m\over \partial t_n}=F_n\frac{\partial Q_m}{\partial
x}+Q_n\frac{\partial F_m}{\partial x}.
\label{eq:divergence}
\end{equation}
Furthermore, the right hand side is a derivative;
\[
{\partial\Q_m\over \partial t_n} = {\partial\Delta(m,n)\over\partial x}\,,
\]

\noindent where

\begin{equation}
\Delta(m,n) =
{\F_{m,u} \F_{n+1,v} + \F_{m,v} \F_{n+1,u} \over (n+1)(m+n+1) }\,.
\label{eq:delta}
\end{equation}

\smallskip

\noindent {\bf Proof}

The first form of the equation (\ref{eq:divergence}) follows straightforwardly
from (\ref{eq:hierarchy}). The second is most easily seen by the introduction
of a potential.

\indent From (\ref{eq:three}) it is easy to change variables to obtain the
Cauchy-Riemann like equations

\begin{eqnarray*}
\frac{\partial\F_n}{\partial\F_m} & = &{\phantom{P}}
\frac{\partial\Q_n}{\partial\Q_m}\\
\frac{\partial\F_n}{\partial\Q_m} & = &P\frac{\partial\Q_n}{\partial\F_m}
\end{eqnarray*}
\bigskip
Introducing a potential
\[
\F_n=\frac{\partial\Delta(m,n)}{\partial
\Q_m}\,,\quad\Q_n=\frac{\partial\Delta(m,n)}{\partial \F_m}
\] the equation (\ref{eq:divergence}) simply becomes
\begin{equation}
{\partial\Q_m\over \partial t_n}=\frac{\partial\Delta(m,n)}{\partial
Q_m}\frac{\partial Q_m}{\partial x}+\frac{\partial\Delta(m,n)}{\partial
F_m}\frac{\partial F_m}{\partial x}=\frac{\partial\Delta(m,n)}{\partial x}
\label{eq:proof}
\end{equation}
A more detailed analysis shows that the potential $\Delta(m,n)$
is given by (\ref{eq:delta}). This makes use of equations
(\ref{eq:one},\ref{eq:two}) and (\ref{eq:three}). In the case $m=1, \F_1=S,\
\Q_1=P$ the potential is simply given by
$\displaystyle{\Delta(1,n)=\frac{\F_{n+1}}{n+1}}
$

\section*{4. The Hamiltonian structure of the hierarchy}

\bigskip

A system (\ref{eq:hydro}) of hydrodynamic type is said to be Hamiltonian if
there exists a Hamiltonian $H=\int dx \, h({\bf u})$ and a Hamiltonian operator

\[
{\hat A}^{ij} = g^{ij} ({\bf u}) {d\phantom{x}\over dx} +
b^{ij}_{\phantom{ij}k} ({\bf u}) u^k_x
\]

\noindent which defines a skew-symmetric Poisson bracket on functionals

\[
\{I,J\} = \int dx \, {\delta I\over \delta u^i(x) } {\hat A}^{ij}
                     {\delta J\over \delta u^j(x) }
\]

\noindent which satisfies the Jacobi identity and which generates the system

\[
u^i_t = \{ u^i(x), H\}\,.
\]

\noindent Dubrovin and Novikov \cite{DubrovinNovikov} proved necessary and
sufficient
conditions for
${\hat A}^{ij}$ to be a Hamiltonian operator in the case when $g^{ij}$ is not
degenerate. These are:

\smallskip

a) ${\bf g}=(g^{ij})^{-1}$ defines a Riemannian metric,

\smallskip

b) $b^{ij}_{\phantom{ij}k} = - g^{is} \Gamma^{j}_{\phantom{j}sk}\,,$ where
$\Gamma^{j}_{\phantom{j}sk}$ is the Christoffel symbol generated by
${\bf g}\,,$

\smallskip

c) the Riemann curvature tensor of $\bf g$ vanishes.

\smallskip

\noindent The system (\ref{eq:hydro}) may then be written as

\begin{equation}
u^i_t = ( g^{is} \nabla_s \nabla_j h ) u^j_x  \label{eq:hamhydro}
\end{equation}

\noindent where $\nabla$ is the covariant derivative generated by ${\bf g}\,.$
Thus to find the Hamiltonian structure for the hierarchy (\ref{eq:hierarchy})
one needs to find the metric and the Hamiltonians.

\bigskip

One approach is to first diagonalise the system, there being the simple
formulae
for the metric coefficients for diagonal systems \cite{Tsarev}:

\[
{\partial_i V_j \over (V_i - V_j)} = \frac{1}{2} \partial_i \log g_{jj}
\]

\noindent where $V^i_j ({\bf u}) = V_j({\bf u}) \delta^i_j$ (no sum).
This diagonalisation is achieved by the transformation:

\begin{eqnarray*}
U & = & \sqrt{u} + \sqrt{v} \,, \\
V & = & \sqrt{u} - \sqrt{v} \,.
\end{eqnarray*}

\noindent While this enables one to find the metric, the formulae are actually
more elegant in the $\{u,v\}$ coordinates.

\bigskip

\noindent {\bf Theorem}

The Hamiltonian structure is given by the zero-curvature metric

\[
{\bf g} = 2 {du\over u} {dv\over v}
\]

\noindent and the Hamiltonians are given by

\[
H_n = \int dx \, h_n({\bf u})
\]

\noindent where $h_n = (n+1)^{-2} \Q_{n+1}\,.$

\smallskip

\noindent {\bf Proof}

The metric is clearly non-degenerate and flat, and the only non-zero
Christoffel symbols are $\Gamma^{1}_{~11}=-u^{-1}$ and
$\Gamma^{2}_{~22}=-v^{-1}\,$ (here $u^1=u$ and $u^2=v\,$). Hence, on expanding
(\ref{eq:hamhydro}) one obtains the following equations for $h_n$

\begin{eqnarray*}
\F_n & = & u v {\partial^2 h_n\over\partial u \partial v} \,, \\
\Q_n & = & {\partial{\phantom{v}}\over\partial v} \big[ v {\partial
h_n\over\partial v} \big] \,, \\
    & = & {\partial{\phantom{u}}\over\partial u} \big[ u {\partial
h_n\over\partial u} \big] \,.
\end{eqnarray*}

\noindent With $h_n = (n+1)^{-2} \Q_{n+1}$ these reduce to identities derived
in section $2$ and $3$. Hence the result.

\bigskip

\bigskip

This show the elegance of the $\{u,v\}$ coordinate system: in terms of
$\{P,S\}$ or $\{r,\theta\}$ the metric is considerably more complicated.

\bigskip

\noindent {\bf Proposition}

The Hamiltonians $H_n$ are time independent and in involution with respect to
the Poisson bracket given by ${\bf g}\,.$

\smallskip

\noindent {\bf Proof}

The result follows from the formulae already derived.

\begin{eqnarray*}
{d H_n\over dt_m} & = & (n+1)^{-2} \int dx \, {\partial \Q_{n+1}\over
\partial t_m}\,,  \\
& = & (n+1)^{-2} \int dx \,{\partial \Delta(n+1,m)\over\partial x}\,,\\
& = & 0 \,,
\end{eqnarray*}

\noindent under suitable boundary conditions. The second results follows from
the use of the same formula:

\begin{eqnarray*}
\{H_m,H_n\} & = & \int dx \, {\partial h_m\over\partial u^i} {\hat A}^{ij}
                             {\partial h_n\over\partial u^j}\,, \\
& = & (n+1)^{-2} \int dx \,
\F_n{\partial\Q_{m+1}\over\partial x}+\F_n {\partial\F_{m+1}\over\partial
x}\,,\\
& = & (n+1)^{-2} \int dx \, {\partial \Delta(m+1,n)\over\partial x}\,. \\
& = & 0 \,.
\end{eqnarray*}

\bigskip

\bigskip

The factor $(n+1)^{-2}$ may, without loss of generality, be absorbed into a
redefinition of the times $t_n\,.$ One therefore obtains the following
generating function for the Hamiltonians:

\[
{\cal H}(\lambda) = \int dx \, {\sl h}(\lambda)
\]

\noindent where

\[
{\sl h}(\lambda) =
{1\over \lambda\sqrt{ (u-v)^2\lambda^2 - 2(u+v)\lambda +1}} -
{1\over\lambda}\,.
\]

\bigskip

\section*{5. Generalizations}

The appearance of Legrendre polynomials suggests that there might be similar
systems with conservation laws given in terms of more general polynomials.
The system

\begin{eqnarray*}
S_t & = & P_x \,, \\
P_t & = & P^{\omega} S_x
\end{eqnarray*}

\noindent may be transformed into the system

\begin{eqnarray*}
u_t & = & (uv)^{\alpha} u v_x \,, \\
v_t & = & (uv)^{\alpha} v u_x
\end{eqnarray*}

\noindent where $ \alpha=(\omega-1)/(2-\omega)\,$ (the case $\omega=2$ having
to be treated separately). By an analogous proceedure to that outlined above,
one can show that this has conservation laws expressible in terms of Gegenbauer
polynomials $T^\alpha_n\,,$ so when $\alpha=0$ these collapse to Legendre
polynomials. The Hamiltonian structure is also similar; this is given by the
metric

\[
{\bf g} = 2 {du\over u^{\alpha+1} } {dv \over v^{\alpha+1} } \,,
\]

\noindent and the Hamiltonians are directly related to the charges, as before.

\bigskip

More generally still, one can look for polynomial conservation laws for the
system

\begin{eqnarray*}
r_t & = & r^a q(z) z_x \,, \\
z_t & = & r^b p(z) r_x \,,
\end{eqnarray*}

\noindent this being a direct generalisation of equation (\ref{eq:RZtoda}).
Looking for charge densities and fluxes of the form $\F=r^c w(z)\,,\Q=r^d y(z)$
results in the constraints $a-b=2$ and $c=b+d-1\,,$ and the self-adjoint
differential equation

\begin{equation}
{d{\phantom{z}}\over dz} \Big( p(z) {dy \over dz} \Big) - d(d+a-1) y = 0
\label{eq:hyper}
\end{equation}

\noindent for $y(z)\,.$ Assuming that this has regular points at $0\,,1$ and
$\infty$ results in a hypergeometric equation and constraints on the otherwise
free functions $p(z)$ and $q(z)\,.$ To get polynomials solutions requires $d$
to be a non-negative integer, and so finally the charge densities are

\[
\Q_n = r^n\,\, {_2 F_1} (-n,n+a-1,c;z)\,,
\]

\noindent where ${_2 F_1}$ is the hypergeometric series. In terms of the
${u,v}$
variables one obtains:

\begin{eqnarray*}
u_t & = & u^{a-c} v^{c-1} v_x \,, \\
v_t & = & u^{a-c-1} v^{c} u_x \,.
\end{eqnarray*}

\noindent There is, however, considerable degeneracy in this system, with
apparently different systems being connected via a change in variable.
Other systems may be obtained, for example, by assuming that (\ref{eq:hyper})
is a confluent hypergeometric function, and this contains a family of charges
given by Laguerre polynomials.

\section*{Acknowledgements}

We would like to thank Chris Athorne for suggesting the introduction of a
potential in the proof of
the lemma in section 3. I.A.B.S. would like to thank the University of
Newcastle for a
Wilfred Hall Fellowship.

\section*{Appendix}

The results in this paper have been obtained by direct calculation; no
use has been made of any Lax equation. In this appendix the Lax equation will
be studied, and the solution to the associated linear problem derived. The
system (\ref{eq:SPtoda}) is a special case of the more general equation

\[
\partial_t \partial_{t'} \log P = \partial^2_x P \,,
\]

\noindent known as the Boyer-Finley or $SU(\infty)$ Toda equation. This is
itself a reduction of the equations for an anti-self-dual Ricci-flat metric.
Physically the system (\ref{eq:SPtoda}) describes anti-self-dual Ricci-flat
metrics with two commuting Killing vectors, and so any Lax pair will fall into
the class studies by Tod and Ward \cite{TodWard}. One possible Lax pair
for the Boyer-Finley
equation (there are other, equivalent, formulations \cite{TakasakiTakebe}) was
given by Ward
\cite{Ward}, and
imposing the symmetry $\partial_t = \partial_{t'}$ gives a Lax pair for
(\ref{eq:SPtoda}).

\bigskip

Explicitly, a Lax pair for the dispersionless Toda system (\ref{eq:SPtoda}) is
given by

\begin{eqnarray*}
{\cal L}_0 & = & \partial_t - L_f - \xi L_{e^{+}} \,, \\
{\cal L}_1 & = & \xi (\partial_t + L_f ) + L_{e^{-}} \,,
\end{eqnarray*}

\noindent where $[L_f,L_g] = L_{ \{f,g\} }$ with $\{f,g\}=f_y g_x - f_x g_y$
and $e^{\pm} = P^{1\over 2} (x,t) \exp (\pm y) \,,$ $f={1\over 2} S(x,t)\,.$
Then the integrability condition $[{\cal L}_0,{\cal L}_1]=0$ for this otherwise
overdetermined linear system

\begin{eqnarray*}
{\cal L}_0 \Psi & = & 0 \,, \\
{\cal L}_1 \Psi & = & 0 \,,
\end{eqnarray*}

\noindent gives, on equating the coefficients of the various powers of
$\xi\,,$ the spectral parameter, to zero, the system $P_t = P S_x\,,$
$S_t = P_x\,.$

\bigskip

The solution $\Psi$ to the linear system may be found in the form
of a power series

\[
\Psi = \sum_{n=0}^{\infty} \psi_n(x,t) \exp ( ny) \xi^n\,.
\]

\noindent The $\psi_n$ are not unique -- one can multiple by an arbitrary
holomorphic function of $\xi$ and this will change all the coefficients, but
not the integrability conditions, this corresponding to a gauge freedom.
This yields an infinite number of recursion
relations

\begin{eqnarray*}
\partial_x \theta_{n+1} & = & P^n \partial_t ( P^{-n} \theta_n ) \,,      \\
\partial_t \theta_{n+1} & = & P^{n+1} \partial_x ( P^{-n} \theta_n ) \,,
\end{eqnarray*}

\noindent where

\[
\theta_n(x,t) = \psi_n P^{n\over 2}\,.
\]

\noindent together with the initial conditions
$\partial_x \theta_0 = \partial_t \theta_1 = 0 \,.$
The first few solutions to these equations are

\begin{eqnarray*}
\theta_0 & = & \phantom{-}k_0 \,,                            \\
\theta_1 & = & \phantom{-}k_1 \,,                            \\
\theta_2 & = & -k_1 S + k_2 \,,                   \\
\theta_3 & = & \phantom{-}k_1 (S^2 - P) - 2 k_2 S + k_3\,.
\end{eqnarray*}

\noindent Here the $k_i$ are constants. Using the above-mentioned gauge freedom
one can set $k_0 = 0$ and $k_i = 0$ for $i \geq 2\,,$ so that, for $n\geq 1\,,$
$\theta_n$ is an $(n-1)^{\rm th}$ polynominal in $u$ and $v\,.$
After a few more iterations one finds the relation

\[
\theta_{n+1} + S \theta_n + P \theta_{n-1} = 0 \,,\quad n\geq 2\,
\]

\noindent appears to work, and this can be proved by induction (again using the
the gauge freedom to set various constants to zero). Solving this relation
gives

\[
\theta_n =  (-1)^{n+1} k_1 \Bigg( {u^n - v^n \over u - v} \Bigg)
\]

\noindent and hence $\psi_n\,.$ Summing the power series is straightforward:

\[
\Psi = {k_1\over \sqrt{uv} }\,\, {1\over {{\xi}}^{-1}\exp(-y) +
\Big[ \sqrt{ u\over v} + \sqrt{ v\over u } \Big] + {{\xi}}\exp(+y) }
\]

\noindent where $\tilde{\xi} = \xi \exp{y}\,.$ In all this $y$ plays
an auxiliary r{\^o}le. This may be removed in favour of
$\tilde{\xi}=\xi\exp(+y)$,
but this will introduce terms involving $\partial_{\tilde{\xi}}\,$ in the
Lax pair.


\bigskip

\begin{thebibliography}{~~}

\bibitem{Ward} R.S.Ward, {\sl Class. Quantum Grav.} {\bf 7} (1990) L95,
{\sl J. Geom. Phys.} {\bf 8} (1992) 317.

\bibitem{Krichever} I.M.Krichever, {\sl Commun. Pure Applied Math.} {\bf 47}
(1994) 437.

\bibitem{Saveliev} M.Saveliev, {\sl On the integrability problem of the
continuous long wave approximation of the Toda lattice}, preprint ENSL, Lyon,
1992.


\bibitem{Hoppe}J. Hoppe and Q-Han Park, Infinite Charge Algebra of
Gravitational Instantons, {\sl Phys. Lett.} {\bf B321} (1994) 333-337.

\bibitem{min}M. B. Mineev-Weinstein, Conservation Laws in Field Dynamics or Why
Boundary Motion is Exactly Integrable? Los Alamos preprint Ms-B258 (1995)

\bibitem{DubrovinNovikov} B.A.Dubrovin and S.P.Novikov, {\sl Soviet Math.
Doklady} {\bf 27} (1983) 665.

\bibitem{Tsarev} S.P.Tsarev, {\sl Soviet Math. Doklady} {\bf 31} (1985) 488,
{\sl Math. USSR Izves.} {\bf 37} (1991) 397.

\bibitem{Kodama} Y.Kodama, {\sl Phys. Lett.} {\bf A147} (1990) 477.

\bibitem{FairlieMulvey} D.B.Fairlie and J.A.Mulvey, {\sl J. Phys.} {\bf A27}
(1994) 1317.

\bibitem{Olver} P.J.Olver, {\sl Applications of Lie Groups to Differential
Equations}, (Springer, New York, 1986).

\bibitem{TodWard} K.P.Tod and R.S.Ward, {\sl Proc. R. Soc.} {\bf A386} (1979)
411.

\bibitem{TakasakiTakebe} K.Takasaki and T.Takebe, {\sl Lett. Math. Phys.}
{\bf 23} (1991) 205.

\end{thebibliography}
\end{document}